\documentclass[aps,prd,twocolumn,superscriptaddress,amsfonts,amssymb,amsmath,eqsecnum,nofootinbib,floatfix,showpacs]{revtex4}

\usepackage{graphicx,color}
\definecolor{darkblue}{rgb}{0,0,0.7}
\definecolor{darkred}{rgb}{0.7,0,0}
\usepackage[unicode, colorlinks, citecolor=darkblue, linkcolor=darkred, urlcolor=blue]{hyperref}

\allowdisplaybreaks

\begin{document}
\date{\today}
\title{Highly reflective low-noise etalon-based meta-mirror}
\author{Johannes~Dickmann}
\email{johannes.dickmann@ptb.de}
\affiliation{Physikalisch-Technische Bundesanstalt, Bundesallee 100, 38116 Braunschweig, Germany}

\author{Stefanie~Kroker}
\affiliation{Physikalisch-Technische Bundesanstalt, Bundesallee 100, 38116 Braunschweig, Germany}
\affiliation{Technische Universit\"at Braunschweig, LENA Laboratory for Emerging Nanometrology, Pockelsstra{\ss}e 14, 38106 Braunschweig, Germany}

\begin{abstract}
We present a concept of a mirror for the application in high-reflectivity low-noise instruments such as interferometers. The concept is based on an etalon with a metasurface (meta-etalon) on the front and a conventional multilayer stack on the rear surface. The etalon in combination with the metasurface enables a dedicated spatial weighing of the relevant thermal noise processes and by this a substantial reduction of the overall read out thermal noise. We exemplary illustrate the benefit of the proposed etalon for thermal noise in two applications: The test masses of the Einstein Telescope gravitational wave detector and a single-crystalline cavity for laser frequency stabilization. In the Einstein Telescope the thermal noise of the etalon even at room temperature outperforms existing concepts for operation temperatures at 10\,K. For the laser stabilization cavity, a reduction of the modified Allan deviation of an order of magnitude is predicted.
\end{abstract}

\maketitle

\section{Introduction}
\label{sec:intro}

Thermal noise limits the sensitivity of high-precision measurement devices like interferometric gravitational wave detectors and Fabry-P\'erot cavities for the frequency stabilization of lasers \cite{harry2010advanced, virgo, kagra, kessler2012sub, audley2017laser}. Among the noise sources, Brownian thermal displacement noise of the optical mirror coatings sets the most severe limitation. There are two approaches to reduce thermal noise: The first one addresses the material properties of the optical coatings in stacks of alternating dielectric layer pairs. In this case, the coating thermal noise is mainly determined by the mechanical loss of the coating layers \cite{harry2002thermal, principe2015,granata2016,abernathy2018}. The mechanical loss can substantially be reduced by the use of crystalline coating layers instead of amorphous materials \cite{cole2013tenfold,cole2016high, cumming2015measurement,lin2015}. However, thermal noise scales with the coating thickness. Equally, the reflectivity of multilayer stacks increases with the number of layer pairs. For example, coating stacks with a reflectivity of $>99.9994\%$ typically require 30 to 40 quarter-wavelength layers \cite{harry2010advanced,abernathy2011einstein,acernese2014advanced}. This inherent relationship between reflectivity and thermal noise sets a limit to the ultimate noise performance that can be achieved.

The second approach waives the use of alternating layer-pairs and therewith the increase of thermal noise with reflectivity. The approach is based on periodic sub-wavelength structures (hereinafter: metasurface) manufactured from a dielectric material with high refractive index. These structures are designed to provide an optical resonance based on two coupled Bloch-modes \cite{lalanne,karagodsky2010theoretical}. The modes interfere constructively in the backward direction of the incoming light enabling a high reflectivity with high spectral and angular tolerance \cite{brueckner, krokerol}. The Bloch-modes are localized in a surface layer with a thickness of less the wavelength of light \cite{karagodsky2010theoretical}. With respect to thermal noise the main advantage of the metasurfaces is, that their reflectivity does not scale with their thickness. The minimum thickness is about $\lambda/4n$, where $\lambda$ is the wavelength of light and $n$ the refractive index of the metasurface. Basic proof-of-principle experiments in interferometry have been performed \cite{friedrich}. However, the maximum reflectivity of $99.8\%$ that was experimentally achieved so far is not sufficient \cite{brueckner} for application in interferometric gravitational wave detectors or laser cavities for frequency stabilization \cite{kessler2012sub,abernathy2011einstein}.

In the past years, advances for the rigorous computation of thermal noise of arbitrarily shaped reflective surfaces were reached \cite{heinert2013calculation, kroker2017brownian,dickmann2018influence}. This lays the foundation for a deeper understanding of the complex interplay of dissipative processes and thermal noise in these systems and prepares the ground for new possibilities in the design of low-noise optical elements.

In this contribution we present a concept which overcomes limitations in the reflectivity of metasurfaces and simultaneously provides an excellent thermal noise performance. This is realized by combining the optical functions of a conventional multilayer mirror and a metasurface while suppressing the coupling of mechanical fluctuations between each other by the use of an anti-resonant etalon (hereinafter: meta-etalon). To illustrate the potential of the proposed concept, we perform a holistic analysis of thermal noise for the low-frequency detector of the Einstein Telescope gravitational wave detector (ET-LF) \cite{abernathy2011einstein} and for crystalline silicon cavities for the frequency stabilization of laser light \cite{kessler2012sub}.

The article is organized as follows: In Sec.~\ref{sec:optics} we describe the basic layout of the meta-etalon.
In Sec.~\ref{sec:general}, we give a brief introduction into the computation of the relevant thermal noise contributions. In Sec.~\ref{sec:results}, the results of the optical optimization and the noise evaluation are presented and discussed. Computational details and material parameters can be found in the appendix. 

\section{Etalon-based meta-mirror} \label{sec:optics}

\begin{figure}[b]
	\centering
		\includegraphics[width=0.48\textwidth]{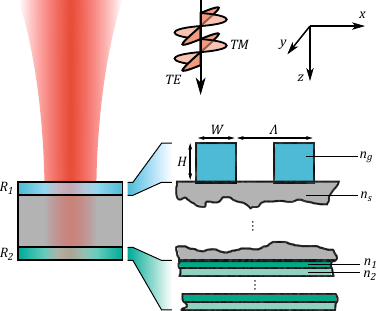}
	\caption{Schematic of the etalon-based meta-mirror for low thermal noise and high reflectivity. The gray area indicates the etalon spacer, the blue one the nano-structured front surface (metasurface) and the green one the rear surface consisting of a conventional multilayer stack. The intensity reflectivities are $R_1$ for the metasurface and $R_2$ for the coating stack. $W$ represents the ridge width, $H$ the ridge height and $\Lambda$ the period of the metasurface. The metasurface material’s index of refraction is denoted $n_g$, the substrate’s $n_s$ and the altering coating’s $n_1$ and $n_2$, respectively. The electric field vector of the incident light is illustrated for transverse electric (TE) and transverse magnetic (TM) polarized light.}
	\label{fig:schematic}
\end{figure}

Fig.~\ref{fig:schematic} shows a schematic of the etalon-based meta-mirror (meta-etalon). The major fraction of the incident light is reflected by the metasurface at the front. Its intensity reflectivity $R_1$ is determined by the refractive indices $n_\mathrm{g}$ and $n_\mathrm{s}$ of the involved materials as well as by the structural parameters $W$, $H$ and $\Lambda$. Details on the metasurface design will be discussed in detail in Sec.~\ref{sec:grating}. The residual light being transmitted by the metasurface propagates through the etalon and is reflected by a conventional multilayer stack (refractive indices $n_\mathrm{1}$ and $n_\mathrm{2}$) on the rear surface of the etalon (intensity reflectivity $R_2$). To achieve a high reflectivity with the whole system, the etalon, forming a two-mirror system, is thermally tuned to anti-resonance \cite{bornwolf}. For both investigated applications, i.e. the Einstein Telescope gravitational wave detector (ET) and the crystalline silicon cavity for laser frequency stabilization \cite{kessler2012sub,matei20171} we choose the etalon spacer material to be fused silica. The metasurface shall be made of single-crystalline silicon and the coating stack on the backside of the etalon shall consist of altering amorphous silica (SiO$_2$) and tantala (Ta$_2$O$_5$) layers. Details on the design can be found in Appendix~\ref{sec:tables}. The influence of alternative spacer materials is discussed in Sec.~\ref{sec:results}. The Einstein Telescope is examined at room temperature, whereas the crystalline silicon cavity with etalon based meta-mirrors is investigated at 124\,K because the linear thermal expansion coefficient vanishes at this temperature being beneficial for the frequency stability.

\section{Thermal noise analysis}\label{sec:general}

In this section, we present basics and assumptions on the thermal noise analysis for the etalon system. 
First, the Brownian noise resulting from thermally activated local transitions between the minima of asymmetric bistable potentials, associated to quasi-degenerate bond states is studied. It leads to a spatially fluctuating surface. The second source of noise are volume fluctuation of the solid, which lead to spatially fluctuating light paths and thus to a fluctuating phase. This noise type is called thermo-elastic (TE) noise. The third noise type - the thermo-refractive (TR) noise - results from fluctuations of the refractive index. For each component of the etalon system, i.e. the metasurface at the front, the etalon substrate and the multilayer stack on the rear surface Brownian, TE and TR noise are investigated. As shown by Evans \textit{et al.}, correlations between TE and TR noise enable a partial compensation of both noise sources summed up to thermo-optic noise \cite{evans2008thermo}. Here, we consider a worst-case scenario without any correlation between the individual noise contributions. This uncorrelated sum of all noise sources provides an upper limit of the overall noise that has to be expected.

We now briefly introduce the physical quantities we use for the discussion of thermal noise. Starting point for both systems, ET and the crystalline silicon cavity, is the determination of the \emph{thermal noise displacement power spectral density} $S$ in m$^2$~Hz$^{-1}$. From that, the mirror thermal noise of the test masses in ET is calculated as \emph{thermal noise displacement spectral density} $\sqrt{S}$. As uncorrelated sum this quantity reads:

\begin{align}
	\sqrt{S} = \left( \sum_i{S_i} \right)^{1/2}.
\end{align}

\noindent The summation includes all noise contributions $S_i$. 
To describe the frequency stability of the silicon cavity in dependence of the integration time $\tau$ \cite{kessler2012sub} instead of thermal displacement noise, the modified Allan deviation $\sigma_y$ is determined. To this end, first the thermal noise power spectral density $S$ is converted into a frequency noise spectral density $\tilde{S}$ in Hz$^2$ Hz$^{-1}$:

\begin{align}
	\tilde{S} = \frac{c^2}{(\lambda L)^2} S.
\end{align}

\noindent Here $c$ is the speed of light in vacuum, $\lambda$ the wavelength of light and $L$ the length of the cavity, respectively. Again, $\tilde{S}$ contains the uncorrelated sum of all noise contributions of the system. From $\tilde{S}$ the modified Allan deviation at the readout frequency $f$ is computed \cite{dawkins2007considerations}:

\begin{align}
	\sigma_y(\tau) = \left(\frac{2\lambda^2}{c^2}\int_0^\infty \tilde{S}(f)\frac{\sin^4{(\pi\tau f)}}{(\pi \tau f)^2} df\right)^{1/2}.
\end{align}

\subsection{Brownian thermal noise}\label{sec:brownian}

In this section we outline the computation scheme of Brownian thermal noise for the etalon from first principles \cite{tugolukov2018thermal,kroker2017brownian}. The scheme employs Levin's approach of virtual pressures \cite{levin1998internal} and the ponderomotive pressures of the light field resulting from Maxwell stress tensor. Starting point is the fluctuation-dissipation theorem (FDT). It relates the dissipated power under the effect of a virtual pressure to the thermal noise of an optical element \cite{levin1998internal}:

\begin{align}\label{FDT}
	S = \frac{2 k_B T}{\pi^2 f^2} \frac{W_\mathrm{diss}}{F_0^2},
\end{align}

\noindent where $k_B$ is the Boltzmann constant, $T$ the temperature of the optical element, $f$ the mechanical readout frequency, $W_\mathrm{diss}$ the dissipated power under the virtual pressure and $F_0$ the surface integral of the virtual pressure. The spatial weighing of the virtual pressure is given by the Maxwell stress tensor (in SI units):

\begin{align}
	\sigma_{ij} = \epsilon_0\epsilon_r E_i E_j +\frac{1}{\mu_0\mu_r} B_i B_j - \frac{1}{2}\left(\epsilon_0 \epsilon_r E^2 + \frac{1}{\mu_0\mu_r}B^2\right)\delta_{ij}. 
\end{align}

\noindent The indices $i$ and $j$ denote the spatial coordinate basis, $E_i$ are the components of the electric field, $B_i$ the components of the magnetic field, $\epsilon_0\epsilon_r$ is the permittivity, $\mu_0\mu_r$ is the permeability and $\delta_{ij}$ is the Kronecker symbol, respectively. The difference of the stress tensor in- and outside the investigated optical surface leads to the ponderomotive light pressure:

\begin{align}
	\vec{p} (\vec{r}) = \Delta \hat{\sigma} (\vec{r}) \frac{\vec{A}}{A}(\vec{r}).
\end{align}

\noindent The quantity $\Delta\hat{\sigma}$ must be evaluated \emph{directly} at the surface as difference between the stress tensor components in vacuum and material, respectively. $\vec{A}/A$ represents the normalized normal vector on the surface. Generally, this expression is evaluated using the transition conditions for electric and magnetic fields at the surface. For example, if the normal vector of the surface is parallel to the direction $i$, and the material is not magnetically active, i.e. $\mu_r=1$, the pressure can be expressed by the continuous field components either inside or outside the material:

\begin{align}
p_i = \frac{1}{2} \left( \frac{D_i^2}{\epsilon_0}\left( \frac{1}{\epsilon_1} - \frac{1}{\epsilon_2} \right) + \epsilon_0 (E_j^2 + E_k^2) (\epsilon_2 - \epsilon_1) \right),
\end{align}

\noindent with the electric displacement field $D_i = \epsilon_0\epsilon E$ and the permittivity $\epsilon_0\epsilon_1$ outside and $\epsilon_0\epsilon_2$ inside the material, respectively.
Applying the ponderomotive pressure, modulated by the readout frequency $f$, on the optical component introduces an elastic deformation energy density $\mathcal E_\mathrm{el}$ into the component. This leads to a dissipation of energy, proportional to the mechanical loss angle $\Phi$ of the material:

\begin{align}\label{eq:Wdiss}
	W_\mathrm{diss} = 2\pi f \int_V \Phi \mathcal E_\mathrm{el} dV.
\end{align}

\noindent With the dissipated energy $W_\mathrm{diss}$ and Eq. (\ref{FDT}) Brownian thermal noise can be computed.
Eq.~(\ref{eq:Wdiss}) illustrates that thermal noise is affected by the spatial distribution of the mechanical loss \textit{in combination} with the spatial distribution of the electromagnetic fields determining the fluctuation readout. 
As a rule of thumb, the detrimental effect of mechanical losses on Brownian noise is the smaller, the lower the amplitudes of the relevant electromagnetic field components are at the surface of the lossy material. That means, to mitigate Brownian thermal noise it is essential to deliberately shape the spatial distribution of the mechanical losses and, if possible, the distribution of the electromagnetic field. This is the reason why highly reflective metasurfaces made of high refractive index materials can exhibit an extraordinary low thermal noise \cite{kroker2017brownian, dickmann2018influence}. In these structures the electromagnetic field is localized within the volume and the field at the surface is reduced (see Sec. \ref{sec:grating}). This leads to a drastically reduced, i.e. optimized, readout of thermal noise. The computation of thermal noise in binary high-reflectivity metasurfaces is discussed in detail in \cite{dickmann2018influence}.

Brownian thermal noise of plane etalons coated with multilayer mirrors on front and back surface can be described by the approach investigated in \cite{somiya2011reduction, gurkovsky2011reducing}. However, in the present case the periodic metasurface requires to additionally apply a spatially oscillating pressure to the etalon substrate. The spatial oscillation period is the period of the metasurface. Thus, additionally the elastic energy due to this oscillation $\mathcal E_\mathrm{osc}$ must be considered. For this problem an analytical solution does not exist yet. A comprehensive numerical computation with finite element analysis is challenging, because the spatial oscillating period is much smaller than the whole etalon system (e.g. in the case of ET by a factor of $500\, 000$ smaller), which would inevitably lead to an immense number of vertexes in the finite element simulation. To circumvent this problem, we develop a semi-analytical approach. We analytically calculate the deformation energy density induced by the smooth part of the pressure, which is determined by the shape of the incident light beam. The spatial intensity profile of a Gaussian distributed light beam reads:

\begin{align}
	I(r)=I_0 \exp\left( -\frac{r^2}{r_0^2} \right),
\end{align}

\noindent where $r$ is the distance from the center of the beam, measured on the reflective surface. $I_0$ represents the intensity at the center $r=0$ and $r_0$ is the Gaussian beam radius, where the intensity has dropped to $1/e\ I_0$. Not the total ponderomotive pressure acts on the front surface of the etalon, because a small fraction $1-R_1$ of the light is transmitted. Therefore, the pressures are scaled with the coefficients $e_1$ for the front surface and $e_2$ for the rear surface, respectively. For an etalon tuned to anti-resonance, the coefficients can be expressed by \cite{gurkovsky2011reducing}:

\begin{align}
		\label{eq:e2}
		e_1 &= \frac{\sqrt{R_1}\left[ 1+(1+n_s)\sqrt{R_1R_2} +R_2 \right] +\sqrt{R_2}(1-n_s) }{(1+\sqrt{R_1R_2})^2}, \\
		e_2 &= \frac{n_s \sqrt{R_2}(1-R_1)}{(1+\sqrt{R_1R_2})^2}.
		\label{eq:e1}
\end{align}

\noindent Thus, by adjusting the reflectivities of the etalon's front and back side it is again possible to minimize the thermal noise readout.

The spatially oscillating part of the elastic deformation energy is determined for one single period of the metasurface using COMSOL \cite{multiphysics2015v}. The intensity of the incident light is assumed to be constant over the structural period. This is valid if the Gaussian beam diameter is much larger than the period. For ET the period is about $ 6\, 700$ times smaller than the beam radius and for the silicon cavity it is about $500$ times smaller. The numerical result for a single period is then scaled to the whole Gaussian readout following the scheme in \cite{dickmann2018influence}. Both contributions to the elastic energy, the smooth and the oscillating, are then summed to provide an upper limit of the overall noise.
In the limit of a large substrate size with thickness $h\gg r_0$ and diameter $d\gg r_0$ the etalon thermal noise can be computed by (using \cite{bondu1998thermal}):

\begin{align}
		S_\mathrm{B}^\mathrm{sub} = \frac{4 k_B T}{\pi f} \left( \mathcal E_\mathrm{osc} + e_1^2\frac{1}{2\sqrt{2\pi}}\frac{1-\nu^2}{r_0Y} \right) \Phi,
		\label{eq:BSub}
\end{align}

\noindent with the Poisson's ratio of the substrate $\nu$ and the Young's modulus $Y$. $\Phi$ is the loss angle of the substrate material.

The third contribution to Brownian noise results from the coating stack at the backside of the etalon. In the far field, at the backside of substrates with thicknesses much larger than the wavelength, any spatial modulation of the light field caused by the metasurface can be neglected. The etalon we consider here, has a thickness which is by a factor of 4000 larger than the wavelength of light. Thus, the field distribution has in good approximation a Gaussian profile -- both in reflection and in transmission. With this approximation coating thermal noise can be calculated with the model by Nakagawa \textit{et al.} \cite{nakagawa2002thermal}. The respective power spectral density now has to be scaled with the factor $e_2^2$. It reads:

\begin{align}
		S_\mathrm{B}^\mathrm{lay} = e_2^2\frac{2k_BT}{\pi^2 f}\frac{(1+\nu) (1-2\nu)}{Y}\frac{d}{r_0^2} \Phi,
		\label{eq:BCoat}
\end{align}

\noindent where $\nu$ is the mean Poisson's ratio and $Y$ the averaged elastic modulus of the layer stack, respectively. $\Phi$ is the averaged mechanical loss angle of the two coating materials and $d$ represents the total thickness of the layer stack. 

\subsection{Thermo-elastic noise}

A non-linear component of the thermo-optic noise is the thermo-elastic (TE) noise. In contrast to conventional
mirrors, it turned out not to be negligible in the composition of thermal noise of etalon based reflectors.
However, the TE noise is studied very well for conventional mirrors and etalons \cite{evans2008thermo, gurkovsky2011reducing}. The fluctuation-dissipation theorem is applicable to this noise type, as well. The main difference to the computation of Brownian noise is that here the dissipation mechanisms is the heat flow caused by local volume fluctuations. In its general form, this dissipated power is \cite{liu2000thermoelastic}:

\begin{align}
\label{eq:WdissTE}
		W_\mathrm{diss} = 2\pi \kappa T \left( \frac{Y\alpha}{(1-2\nu)C\rho} \right)^2 \int_h\int_0^R \left[\nabla \theta\right]^2rdr,
\end{align}

\noindent where $\kappa$ is the thermal conductivity, $C$ is the thermal capacity per unit volume, $Y$ is the Young's modulus, $\nu$ is the Poisson's ratio, $\alpha$ is the thermal expansion coefficient and $\rho$ is the density, respectively. The quantity $\theta$ is defined as the trace of the strain tensor, which reads in cylindrical coordinates:

\begin{align}
		\theta = \epsilon_{rr} + \epsilon_{\phi\phi} + \epsilon_{zz}.
\end{align}

\noindent The thermo-elastic noise of the substrate and the multilayer on the rear surface can be computed with the dissipated energy given by Eq. (\ref{eq:WdissTE}). In accordance to the work by Heinert \textit{et al.} \cite{heinert2013calculation} the metasurface contribution to thermo-elastic noise is evaluated numerically by means of rigorous coupled wave analysis (RCWA) \cite{moharam1981rigorous}. The computation is performed as follows: Generally, a temperature change $\Delta T$ leads to a relative relative length change $\delta = \alpha \Delta T$. In the case of small temperature changes $\Delta T \ll T$, this results in a linear change of the reflected light’s phase:

\begin{align}
		\delta \varphi = K_\mathrm{TE}\delta,
\end{align}

\noindent where $K_\mathrm{TE}$ is a numerically determined proportional factor. For this computation, the following three contributions must be considered: 1. The geometrical change of the metasurface ridge's height and width, 2. The effective movement of the ridges towards the incident light and 3. The change of the metasurface period. For more details on the implementation of the numerical analysis, see Appendix \ref{sec:calculus}.
Thus, the metasurface's contribution to TE noise reads \cite{heinert2013calculation}:

\begin{align}
		S_\mathrm{TE}^\mathrm{grat} = \left( \frac{\lambda}{4\pi} K_\mathrm{TE}\alpha \right)^2 S_T,
\end{align}

\noindent where $S_T$ represents the noise power of temperature fluctuations introduced in \cite{braginsky2003thermodynamical}:

\begin{align}
		S_T = \frac{k_BT^2}{\pi^{3/2} r_0^2\sqrt{\rho C \kappa f}}.
\end{align}

\subsection{Thermo-refractive noise}

The second component of the thermo-optic noise results from the spatially fluctuating index of refraction in the material crossed by the light field \cite{gurkovsky2011reducing}. For the substrate TR noise, the power spectral density reads \cite{gurkovsky2011reducing}:

\begin{align}
		S_\mathrm{TR}^\mathrm{sub} = e_2^2 \frac{k_BT^2\beta\kappa d}{\pi^2 \rho^2 C^2 r_0^4 f^2} \left( 1+\frac{1}{1+(4\pi/\lambda\sqrt{\kappa/2\pi C\rho f})^4} \right),
		\label{eq:TR_sub}
\end{align}

\noindent where $\beta=dn/dT$ is the thermo-optic coefficient. Similarly to Brownian noise, also the thermorefractive noise scales with the weighing coefficients $e_i$.
The second term of (\ref{eq:TR_sub}) results from the standing wave inside the etalon tuned to anti-resonance. For the coating TR noise follows \cite{gurkovsky2011reducing}:

\begin{align}
\label{eq:TRcoat}
	S_\mathrm{TR}^\mathrm{coat} = e_2^2\times\frac{k_BT^2\beta_\mathrm{eff}^2\lambda_0^2}{\pi^{3/2}\sqrt{\kappa\rho C}r_0^2\sqrt{f}},
\end{align}	
	
\noindent with the effective thermo-optic coefficient of \cite{gurkovsky2011reducing}

\begin{align}	
	\beta_\mathrm{eff}=\frac{1}{4}\frac{\beta_1n_2^2+\beta_2n_1^2}{n_1^2-n_2^2},
\end{align}

\noindent where the index 1 indicates the quantity of material 1, i.e. fused silica and the index 2 indicates the material 2 of the multilayer, i.e. tantala. 
The metasurface contribution to TR noise is again computed using the RCWA code. The computation is performed as follows: A change in the refractive index of the metasurface material $\Delta n$ changes the phase of the reflected light. For small values $\Delta n \ll n$, the phase change of the reflected light is proportional to this change:

\begin{align}	
	\Delta\varphi = K_\mathrm{TR} \Delta n,
\end{align}

\noindent with the proportional constant $K_\mathrm{TR}$. This constant is evaluated numerically (see Appendix \ref{sec:calculus}). Thus, the noise power can be expressed as \cite{heinert2013calculation}

\begin{align}
		S_\mathrm{TR}^\mathrm{grat} = \left( \frac{\lambda}{4\pi} K_\mathrm{TR}\beta \right)^2 S_T.
\end{align}

\section{Results and discussion}\label{sec:results}

\subsection{Optical optimization}\label{sec:grating}

\begin{figure}[b]
	\centering
		\includegraphics[width=0.5\textwidth]{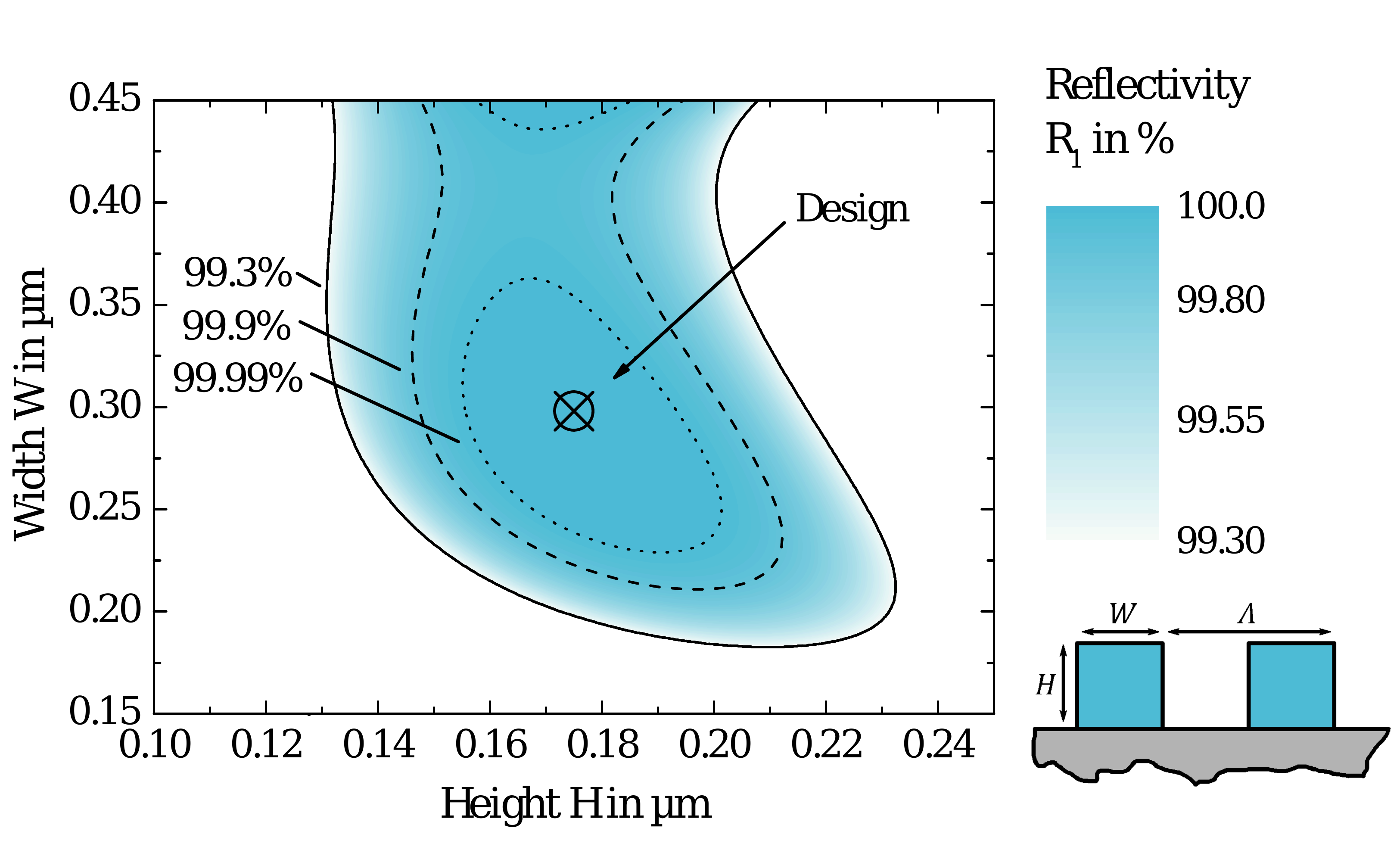}
	\caption{Contour plot of the metasurface reflectivity in dependence of structure height $H$ and width $W$. The isolines indicate the parameter spaces for reflectivities of $99.3\%$, $99.9\%$ and $99.99\%$, respectively.}
	\label{fig:tolerances}
\end{figure}

\begin{table}[b]
	\centering
		\caption{Overview of the optical etalon parameters $e_1$ and $e_2$ for different reflectivities of the metasurface $R_1$.}
		\begin{tabular}{|c|c|c|c|}
		\hline\hline
		Metasurface refl. $R_1$ & Coat. refl. $R_2$ & $e_1$ & $e_2$\\ 
		\hline\hline
		$99.3\%$ & $99.999\,4\%$ & $0.997\,454$ & $0.000\,790$ \\
		$99.9\%$ & $99.999\,4\%$ & $0.999\,637$ & $0.000\,113$ \\
		$99.99\%$ & $99.999\,4\%$ & $0.999\,964$ & $0.000\,011$ \\
		\hline\hline
		\end{tabular}
		\label{tab:e1e2}
\end{table}

The optical design of the etalon configuration aims at a maximum reflectivity in combination with low thermal noise. As demonstrated in the previous section, the noise contributions of front- and backside have to be weighed by the coefficients $e_1$ and $e_2$ which, in turn, depend on the front- and backside reflectivities $R_1$ and $R_2$. Thus, in the etalon configuration Brownian and thermo-optic noise are actually coupled whereby the coupling is dictated by $R_1$ and $R_2$. In this section we illustrate the influence of metasurface reflectivity $R_1$ and the resulting fabrication tolerances on the weighing factors $e_1$ and $e_2$. To this end, the overall transmission of the combined etalon system shall be $<$6\,ppm and we assume a coating stack reflectivity $R_2$ of $99.9994\%$ as a typical value for reflectivities of high-performance multilayer mirrors \cite{harry2010advanced} (see Tab. \ref{tab:Optical}).

\begin{figure}[t]
	\centering
		\includegraphics[width=0.5\textwidth]{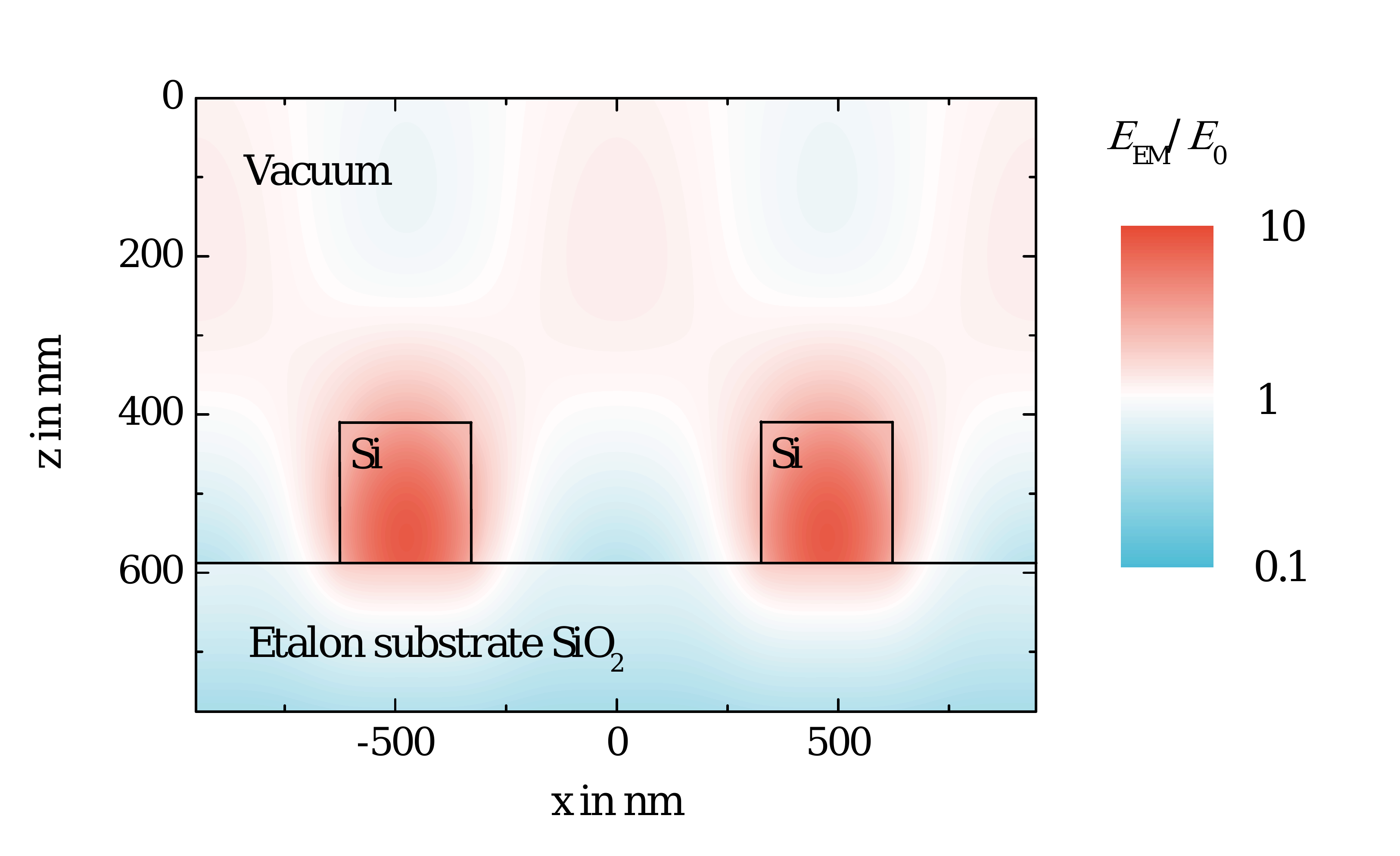}
	\caption{Contour plot of the electromagnetic field energy density in the silicon metasurface of the etalon (metasurface parameters listed in Tab. \ref{tab:Optical}). The energy density is normalized to the incident energy density. The plot shows the spatial distribution in the $x-z$ plane of two unit cells of the structure. The energy density is concentrated in the silicon, but it is only slightly enhanced at the surface of the silicon. In addition, the energy density in the area between the ridges is reduced and inside the etalon substrate the energy density decreases rapidly. This spatially modulated light energy density distribution leads to the minimized readout of thermal noise.}
	\label{fig:emenergy}
\end{figure}

The structure parameters of the metasurface, a subwavelength binary grating structure with one-dimensional periodicity, are determined by means of RCWA \cite{moharam1981rigorous} (wavelength 1.55 $\mu$m). As discussed in Sec.~\ref{sec:brownian}, the use of structures made of a material with high refractive index, e.g. silicon, are promising for low thermal noise. Here, we consider the metasurface to be made of crystalline silicon on a fused silica substrate (silicon on insulator, SOI).
We investigate transverse-electric polarized (TE) light because it is beneficial for the noise performance of metasurfaces with one-dimensional periodicity \cite{dickmann2018influence}. 
Structure period, ridge height and ridge width are optimized to achieve high reflectivities \textit{in combination} with large tolerances for the ridge height and ridge width being the critical parameters in the fabrication process. This parameter range is maximized for a period of $\Lambda=950$\,nm. Fig.~\ref{fig:tolerances} shows the reflectivity $R_1$ in dependence of the metasurfaces' ridge height and width. The permitted fabrication tolerances for three exemplary reflectivities are in the range of a few tens of nanometers. For binary structures these tolerances are realistic in terms of technological feasibility \cite{wurm2017metrology}.\footnote{The metasurfaces can be fabricated, for example, via electron beam lithography and subsequent reactive ion etching.} 

Besides the residual transmission of the metasurface, material absorption in silicon may limit the feasible reflectivity. We evaluated the influence of the silicon material absorption on the intensity absorption of the metasurface by means of RCWA. The measured absorption coefficient of silicon is as low as ${5\times 10^{-6}/\mathrm{cm}}$ \cite{degallaix2013bulk}. To account for a potentially enhanced absorption due to the large surface-to-volume ratio of the metasurface, as a worst-case scenario, we assume the absorption coefficient to be enhanced by a factor of $100$ \cite{khalaidovski2013indication}. In this case the intensity absorption of the metasurface is still smaller than $10^{-10}$ and thus can be neglected.

A typical spatial distribution of the electromagnetic energy density in such a high-reflectivity structure is illustrated in Fig.~\ref{fig:emenergy}. The energy density is concentrated within the volume of the silicon ridges and only slightly enhanced at the surface of the ridges which leads to the aforementioned minimized readout of thermal noise.

Table \ref{tab:e1e2} shows the weighing coefficients $e_1$ and $e_2$ for the three different values of $R_1$. The metasurface's contribution to Brownian thermal noise (scaling with $e_1^2$) does not significantly change with $R_1$. That means $R_1$ is not a critical parameter for Brownian thermal noise. In contrast, the contributions of thermo-refractive noise scaling with $e_2^2$ (compare Eqs.~(\ref{eq:TR_sub}) and (\ref{eq:TRcoat})) change by a factor of more than 5000. Hence, the etalon allows the tuning of the different noise contributions and provides an additional degree of freedom for low-noise high-reflectivity mirrors.

\subsection{Einstein Telescope}

\begin{figure}[t]
	\centering
		\includegraphics[width=0.5\textwidth]{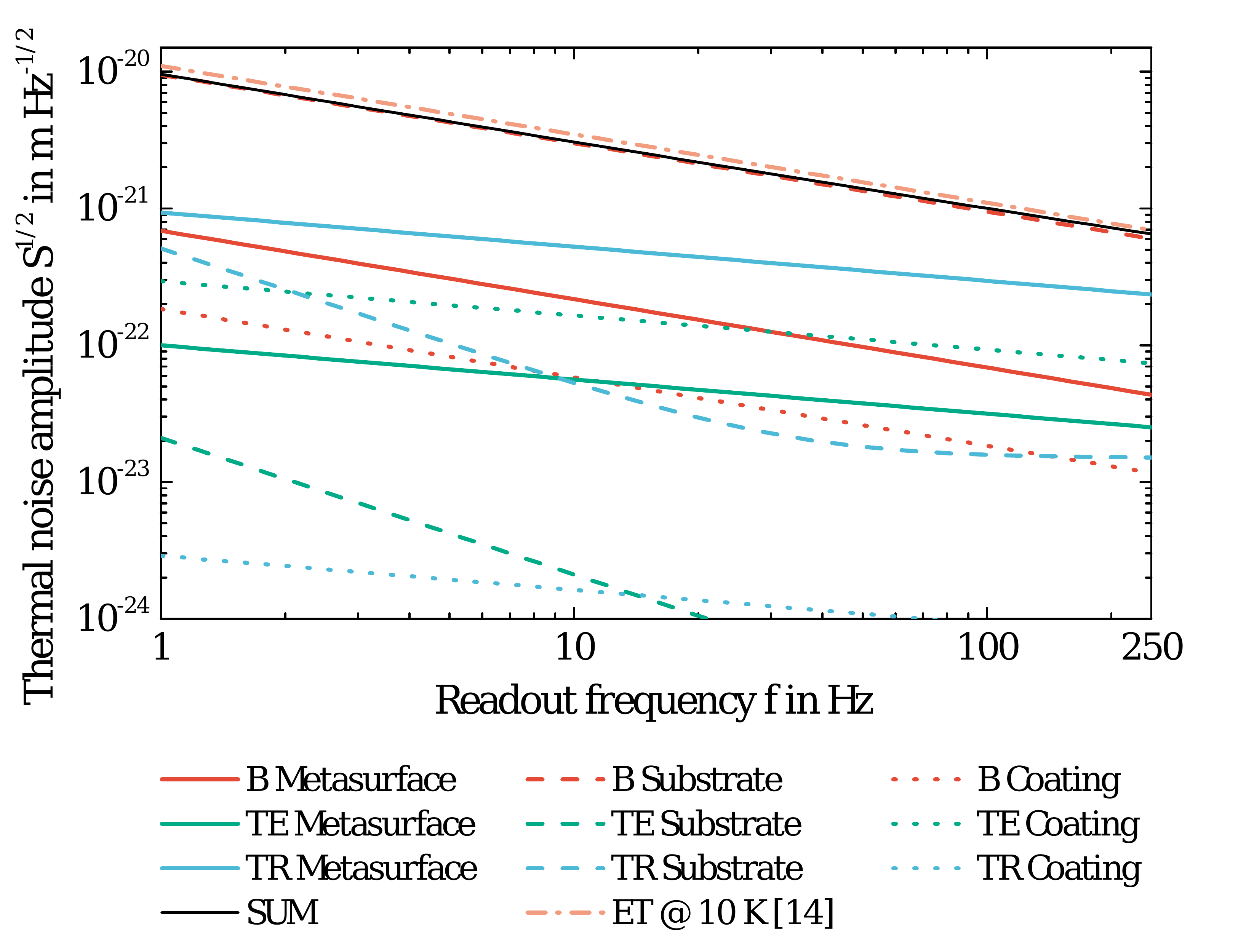}
	\caption{Room temperature thermal noise amplitudes of all meta-etalon components (metasurface, substrate and rear surface coating) for ET-LF versus the mechanical readout frequency. A residual transmission of $0.7\%$ of the metasurface was assumed. The plot shows Brownian (B), thermo-elastic (TE) and thermo-refractive (TR) noise for each component. The black line indicates the uncorrelated sum over all contributions. The current design of ET at a temperature of 10\,K (with state-of-the-art amorphous multilayer mirrors) is shown by the dash-dotted line in magenta color.}
	\label{fig:ET}
\end{figure}

The design parameters for the end mirrors, i.e. the end test masses, of the Einstein Telescope (ET) are summarized in Table \ref{tab:Geometry} at the end of this document. ET is a future interferometric gravitational wave detector and its low frequency part (ET-LF)  shall be optimized for gravitational wave signals with frequencies of about 1 to 250\,Hz \cite{abernathy2011einstein}. To mitigate thermal noise, the mirrors of ET-LF are planned to operate at cryogenic temperatures of about 10 K. The cryogenic operation of ET will entail immense technical effort. The proposed meta-etalon can achieve the cryogenic thermal noise performance of amorphous multilayer mirrors even at room temperature. To illustrate that, we compare thermal noise of multilayer mirrors at a temperature of 10\,K to the meta-etalon at room temperature. All other parameters, such as laser wavelength, arm length and laser power are the same as in the cryogenic design \cite{abernathy2011einstein}. Fig.~\ref{fig:ET} shows the results of the analysis for all thermal noise contributions using a metasurface reflectivity of $R_1=99.3\%$.

In the entire detection bandwidth from 1 Hz to 250\,Hz, thermal noise of the etalon is about 10\% smaller than the sensitivity of ET-LF at 10\,K. Brownian thermal noise of the etalon substrate dominates all other contributions. This means, a further improvement of the metasurface reflectivity $R_1$ is not beneficial, as it only affects thermo-optic (i.e. TE and TR) noise. Compared to mirrors based on standalone high-reflectivity silicon metasurfaces as discussed for ET-LF \cite{brueckner, krokerapl,heinert2013calculation}, meta-etalons enable another improvement of the overall thermal noise. This is thanks to the reduction of Brownian thermal noise as dominant noise source at the expense of thermo-optic noise. The dominance of substrate Brownian noise is additionally remarkable because conventional multilayer mirrors are limited by the Brownian noise of the coating stack. And there are still possibilities for further improvement, especially by using crystalline substrate materials like sapphire or silicon.

\subsection{Single-crystalline silicon cavity} 

The cavities under investigation are made of single-crystalline silicon with a length of 21\,cm. A typical application temperature is the temperature of the zero in the thermal expansion coefficient of silicon at 124 K.

\begin{figure}[t]
	\centering
		\includegraphics[width=0.5\textwidth]{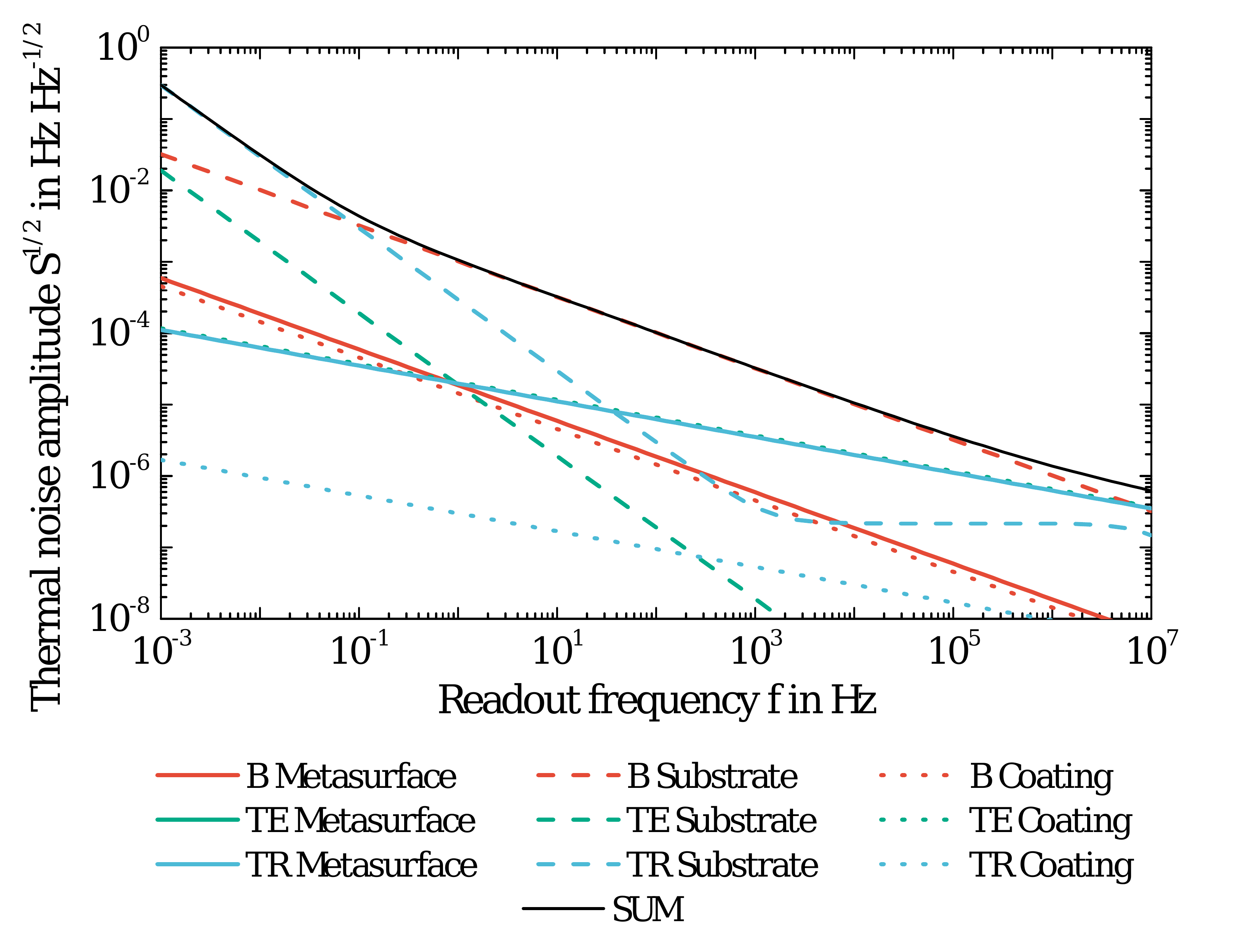}
	\caption{Thermal noise amplitudes of all meta-etalon components (metasurface, substrate and rear surface coating) of the Fabry-P\'erot cavity versus the mechanical readout frequency. A residual transmission of $0.7\%$ of the metasurface was assumed. The diagram shows the Brownian (B), thermo-elastic (TE) and thermo-refractive (TR) noise for each component. The black line indicates the uncorrelated sum over all contributions. The TE coating and TE metasurface noise are smaller than $10^{-8}$\,Hz\,Hz$^{-1/2}$.}
	\label{fig:FP1}
\end{figure}

\begin{figure}
	\centering
		\includegraphics[width=0.5\textwidth]{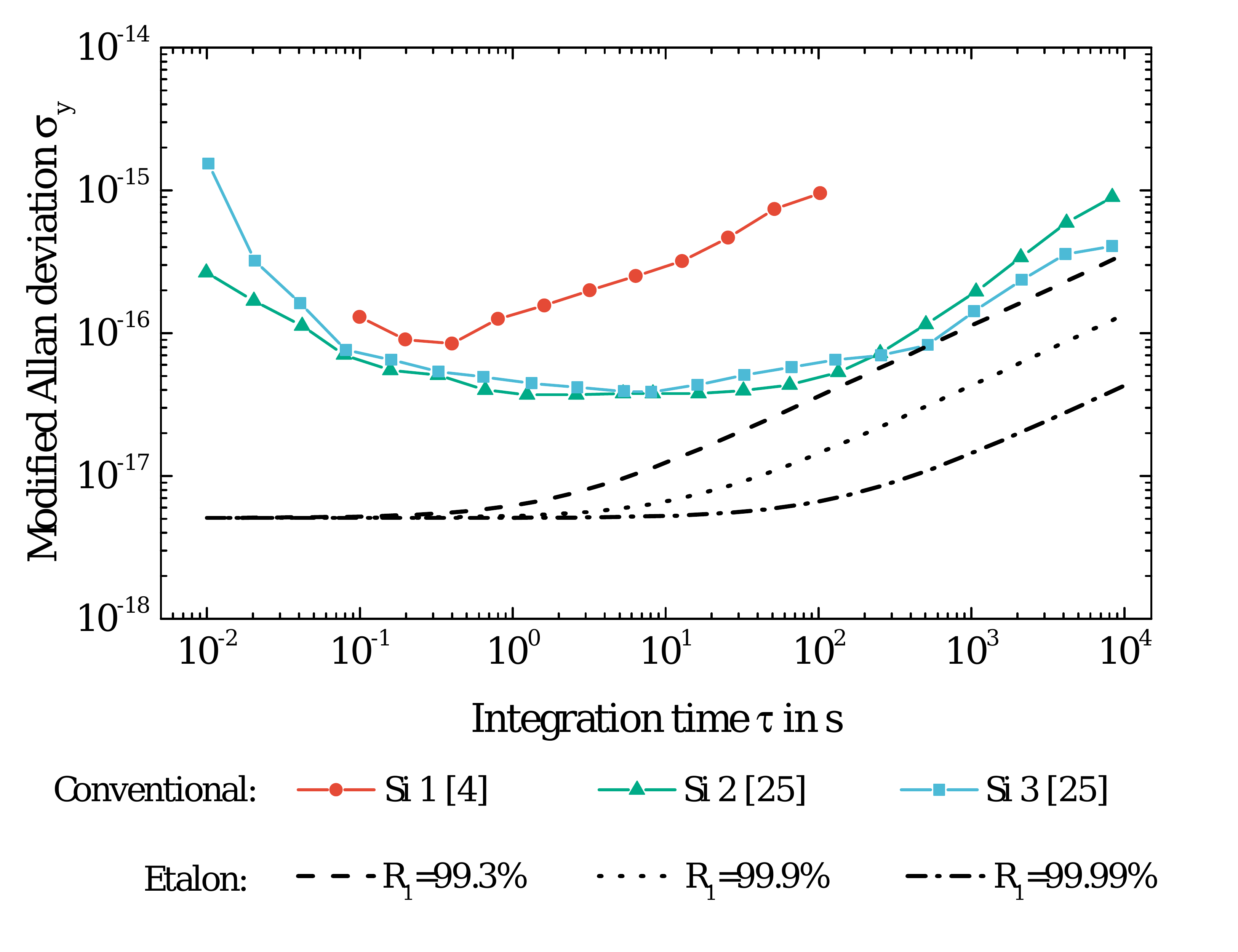}
	\caption{Modified Allan deviation versus integration time for single-crystalline silicon Fabry-P\'erot cavities with different mirrors. Si1-Si3 are cavities with conventional multilayer mirrors. The black lines indicate the cavity performance with meta-etalons as mirrors for different metasurface reflectivities $R_1$.}
	\label{fig:FP2}
\end{figure}

Fig.~\ref{fig:FP1} shows the noise contributions for a silicon cavity with two meta-etalon mirrors for mechanical frequencies from $10^{-3}$ to $10^7$\,Hz. At frequencies $>1$\,Hz Brownian substrate noise dominates. At low frequencies $<0.1$\,Hz the thermo-refractive noise of the substrate makes the main contribution. By changing the residual transmission of the metasurface, the TR contribution of the substrate can be tuned in the following way: For smaller residual metasurface transmissions, the readout of the TR substrate noise decreases with the weighing coefficient $e_2^2$ (compare Sec.~\ref{sec:grating}), because the intensity circulating in the etalon is reduced. Instead, as already mentioned above, the substrate Brownian noise does not change significantly with higher reflectivity of the metasurface, because the effective ponderomotive pressure on the front surface is almost independent of $R_1$. Fig.~\ref{fig:FP2} shows the influence of $R_1$ on the thermal noise as modified Allan deviation for a silicon cavity with two meta-etalons with $R_1=99.3\%$, $99.9\%$ and $99.99\%$. For comparison, the measured stability of the silicon cavities at PTB (Si1-Si3) are illustrated. 

In general, the frequency stability of the meta-etalon based cavity is limited by Brownian substrate noise for small integration times and by the substrate TR noise for large integration times. The integration time, at which TR noise becomes dominant, is crucially affected by the reflectivity $R_1$ of the metasurface. The higher $R_1$, the larger the integration time until which the Brownian noise floor defines the stability. For $R_1=99.99\%$ the stability limitation of the meta-etalon based cavity in terms of thermal noise is about a factor of 10 better than for state-of-the art cavities. The fabrication tolerances of about 15\,nm for structure ridge width and height (see Fig.~\ref{fig:tolerances}) are in reach with available technology. A further improvement of the frequency stability by a factor of up to 100 can be obtained with crystalline materials like sapphire or silicon as etalon substrate or lower temperatures (e.g. a few K).

\section{Conclusion}

In this contribution we presented a concept for a low-noise highly reflective mirror based on an anti-resonant meta-etalon with a metasurface at the front surface and a conventional amorphous multilayer mirror at the back surface. 
To reach a high reflectivity, the meta-etalon must be tuned to optical anti-resonance. This can be done by thermal stabilization.  
Generally, the tolerances for keeping an optical resonator in anti-resonance are large which is promising for a temperature stabilization. In comparison to low-noise metasurfaces with ultra-high reflectivities as standalone mirrors in high-finesse cavities the two-mirror system of the meta-etalon relaxes fabrication tolerances and thus technological challenges in the realization.

We exemplary demonstrated the benefit of the meta-etalon based on a fused silica substrate for two devices: For the Einstein Telescope ET we demonstrated that the etalon achieves the cryogenic noise performance of conventional multilayer mirrors even at room temperature. In crystalline silicon resonators the meta-etalon enables a thermal noise reduction by a factor of 10 with fused silica as substrate material.

The meta-etalon is limited by substrate Brownian noise which can be further reduced by a factor of up to 100 using crystalline substrate materials like sapphire or silicon. In contrast, conventional mirrors are limited by the Brownian thermal noise of the high-reflectivity coating. In the meta-etalon, thermal noise of the meta-surface \textit{and} coating thermal noise are both by a factor of about $10^4$ smaller than substrate Brownian noise and thus negligible. Due to the dedicated spatial weighing of the dissipation processes in the meta-etalon, the use of amorphous coating materials with high mechanical losses of about $10^{-4}$ do not compromise the noise performance of the mirror. This can be considered as a paradigm change for the design and optimization of high precision sensing devices providing new degrees of freedom to optimal optical performance with minimum thermal noise.

\begin{acknowledgments}
The authors thank T. Legero and U. Sterr for the helpful exchange concerning laser stabilization. Furthermore, JD acknowledges A. J. Bergh\"auser for the linguistic support.
\end{acknowledgments}

\appendix

\section{Supplememtal material on thermal noise computation}\label{sec:calculus}

In this appendix we provide supplemental information in the computation of the different  types of thermal noise.

\subsection{Brownian noise}

\subsubsection{Brownian noise---Metasurface}
\label{sec:gratnoise}

The computation of Brownian thermal noise of binary metasurface mirrors is comprehensively discussed in \cite{dickmann2018influence}. The computations base on Finite element simulations with COMSOL Multiphysics \cite{multiphysics2015v} and they are performed over one period of the metasurface structure using Floquet boundary conditions $\vec{k}=2\pi/\Lambda\vec{e}_x$, where $\vec{e}_x$ represents the unit vector in $x$-direction (compare Fig. \ref{fig:schematic} in Section \ref{sec:intro}). The simulation in COMSOL is set up as two-dimensional analysis in the $x$ -- $z$ plane.

In a first step, the electromagnetic field in the structure is computed with a spatially constant energy density of the illuminating light field. The power of the incident light is set to $d\tilde{P}/dy = 1\, \mathrm{W/m}$ (in $y$-direction). Then, the resulting Maxwell stress tensor is applied to the interfaces of the structure. This introduces an elastic deformation energy  of density $d\tilde{E}/dy$ to the metasurface ridges:

\begin{align}
	\frac{d\tilde{E}}{dy} = 5.324\times 10^{-30}\ \frac{\mathrm{J}}{\mathrm{m}}. 
\end{align}

\noindent Following the scheme in \cite{dickmann2018influence}, we evaluate thermal noise for the illumination by the entire Gaussian beam:

\begin{align}
	S_\mathrm{B}^{(1)} = \frac{2k_BT}{\pi^2 f}\frac{\Lambda}{r_0^2}\frac{\mathrm{d}\tilde{E}}{\mathrm{d}y}\left( \frac{2}{c} \frac{\mathrm{d} \tilde{P}}{\mathrm{d} y} \right)^{-2}\Phi.
\end{align}

\subsubsection{Brownian noise---Substrate} As discussed in Section \ref{sec:general}, the substrate Brownian noise consists of two contributions: A smooth contribution due to the Gaussian beam and a spatially oscillating contribution originating from the periodic metasurface. The smooth part is evaluated analytically using Eq. (\ref{eq:BSub}) in Section \ref{sec:general}. The second part is again calculated using a two-dimensional COMSOL simulation. The resulting linear energy density reads:

\begin{align}
	\frac{d\tilde{E}}{dy} = 2.122\times 10^{-30}\ \frac{\mathrm{J}}{\mathrm{m}},
\end{align}

\noindent which is about 4 orders of magnitude smaller than the smooth Gaussian part.

\subsubsection{Brownian noise---Coating stack} The Brownian noise of the coating stack is evaluated analytically using Eq. (\ref{eq:BCoat}) for a coating stack of 18 $\lambda/4$ doublets of silica/ tantala.

\subsection{Thermo-elastic noise}

Substrate and coating thermo-elastic noise are computed analytically following Section \ref{sec:general}. The metasurface's TE noise is evaluated numerically using RCWA \cite{moharam1981rigorous}. Fig. \ref{fig:TEphase} shows the linear relationship between the light phase $\Delta \varphi$ reflected by the metasurface and the parameter $\delta=\alpha\Delta T$. The slope of the line is the coefficient $K_\mathrm{TE}$:

\begin{figure}
	\centering
		\includegraphics[width=0.5\textwidth]{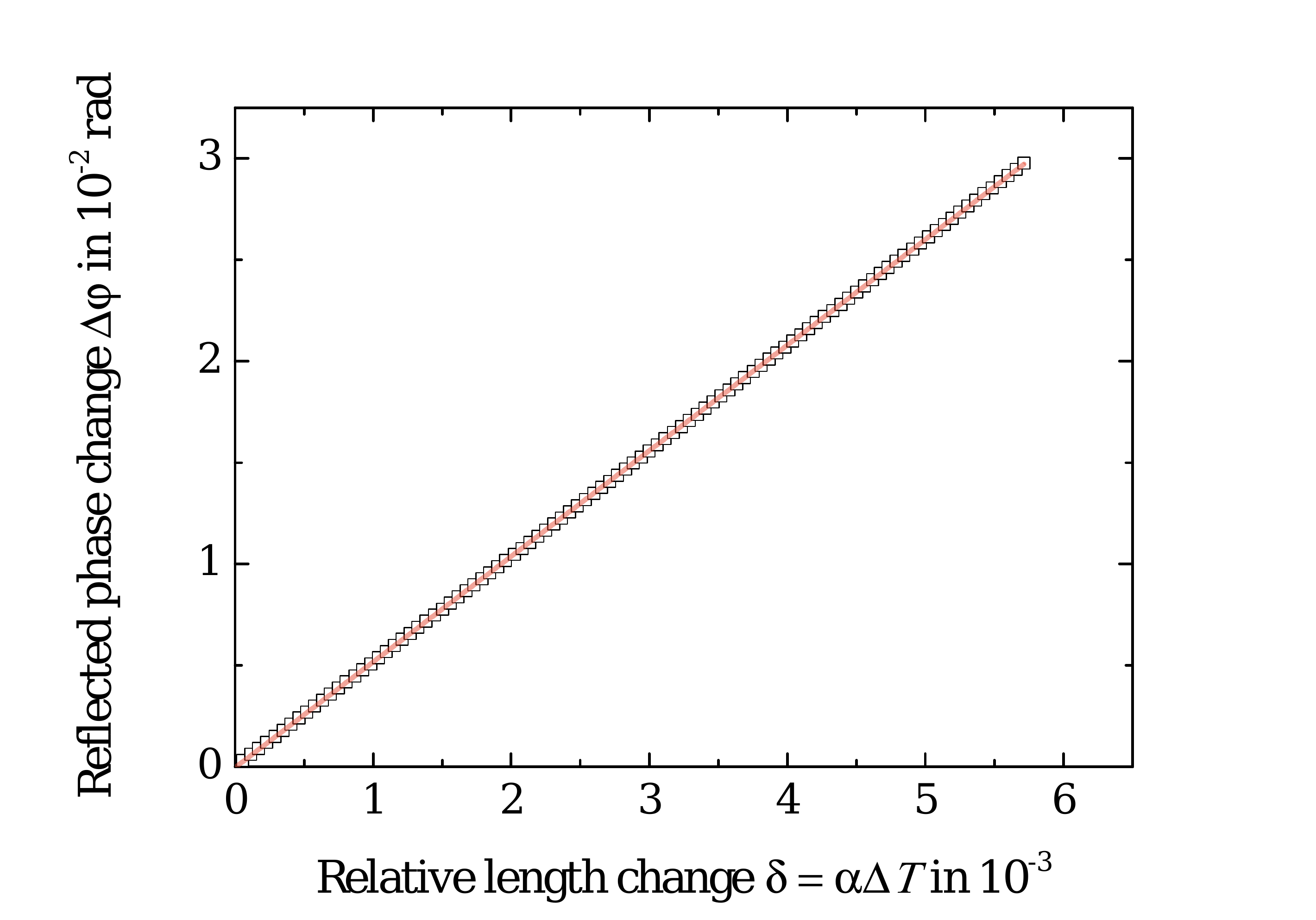}
	\caption{Reflected light phase in dependence of different relative length changes. The values are obtained numerically using RCWA. Red: linear line fit.}
	\label{fig:TEphase}
\end{figure}

\begin{align}
	K_\mathrm{TE} = 5.21.
\end{align}

\subsection{Thermo-refractive noise}

\begin{figure}
	\centering
		\includegraphics[width=0.5\textwidth]{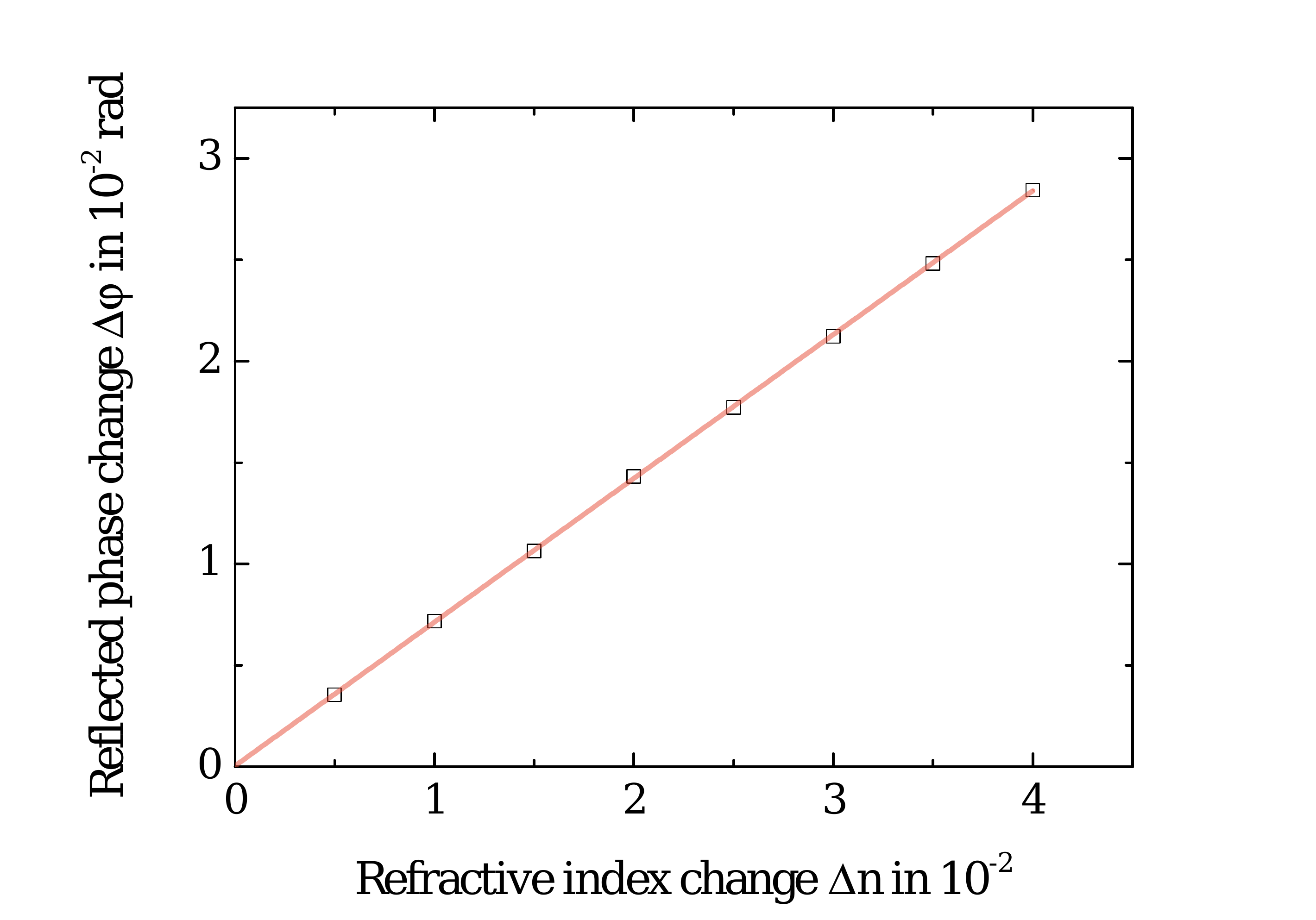}
	\caption{Reflected light phase versus change of the refractive index of the metasurface material (silicon). The values are obtained numerically with RCWA. Red: linear line fit.}
	\label{fig:TRphase}
\end{figure}

Substrate and coating thermo-refractive (TR) noise are evaluated analytically following Section \ref{sec:general}. The parameter for the determination of the metasurface's TR noise, $K_\mathrm{TR}$, is the slope of the plot in Fig. \ref{fig:TRphase}: 

\begin{align}
	K_\mathrm{TR} = 0.71.
\end{align}

\newpage

\section{Tables}\label{sec:tables}

\begin{table}[!htbp]
	\centering
		\caption{Design parameters of the end mirrors of the Einstein Telescope low-frequency detector~\cite{abernathy2011einstein}.}
		\begin{tabular}{|c|c|}
		\hline\hline
		Parameter & Value\\ 
		\hline\hline
		Mirror diameter $d$ & 50 cm\\
		Mirror thickness $h$ & 50 cm \\
		Gaussian beam radius $r_0$ & 6.36 cm \\
		\hline\hline
		\end{tabular}
		\label{tab:Geometry}
\end{table}

\begin{table}[!htbp]
	\centering
	\caption{Parameters of the single-crystalline silicon cavity for the stabilization of laser light at a frequency of $1.55\,\mu$m \cite{matei20171}.}
		\begin{tabular}{|c|c|}
			\hline\hline
			Parameter & Value \\
			\hline\hline
			Cavity length $L$ & 210 mm \\
			Central bore diameter $2a$ & 5 mm \\
			Mirror diameter $d$ & 12.7 mm \\
			Mirror thickness $h$ & 5 mm \\ 
			Gaussian beam radius $r_0$ & 483 $\mu$m \\
			\hline\hline
		\end{tabular}
	\label{tab:parameters}
\end{table}

\begin{table}[!htbp]
	\centering
		\caption{General parameters for the noise computations of the meta-etalon mirror.}
		\begin{tabular}{|c|c|}
		\hline\hline
		 Parameter & Value\\ 
		\hline\hline
		 Wavelength & 1550 nm \\
		  Polarization & TEM$_{00}$ \\ \hline\hline
		 Metasurface material & c-Si\\
		  Metasurface refr. index $n_g$ & 3.48 \\
		  Metasurface period $\Lambda$ & 950 nm \\
		  Metasurface ridge width $W$ & 298 nm \\
		  Metasurface ridge height $H$ & 175 nm \\
		  Metasurf. residual transm. $t_1$ & $<0.7\%$, $<0.1\%$, $<0.01\%$ \\ \hline\hline
		 Substrate material & SiO$_2$ \\
		  Substrate refr. index $n_2$ & 1.45 \\ \hline\hline
		 Coating composition & 18 $\lambda/4$ doublets of SiO$_2$/Ta$_2$O$_5$ \\
		  Coating residual transm. $t_2$ & 6 ppm\\
		\hline\hline
		\end{tabular}
		\label{tab:Optical}
\end{table}

\begin{table}[!htbp]
	\centering
		\caption{Material properties of the meta-etalon components at room temperature~\cite{dickmann2018influence, heinert2013calculation}.}
		\begin{tabular}{|c|c|c|c|c|}
		  \hline\hline
			Parameter & Silicon & Substrate & Ta$_2$O$_5$ layer & SiO$_2$ layer \\
			\hline\hline
			$\beta$, 1/K & $1.8\times 10^{-4}$ & $8\times 10^{-6}$ & $1.4\times 10^{-5}$ & $8\times 10^{-6}$ \\
			$\alpha$, 1/K & $2.62\times 10^{-6}$ & $5.1\times 10^{-7}$ &  $3.6\times 10^{-6}$ & $5.1\times 10^{-7}$ \\
			$\rho$, kg/m$^3$ & $2331$ & $2202$& $6850$ & $2202$ \\
			$Y$, Pa & $130\times 10^9$ & $72\times 10^9$ & $140\times 10^9$ & $72\times 10^9$ \\
			$\nu$ & $0.28$ & $0.17$ & $0.23$ & $0.17$ \\
			$\kappa$, W/(K m) & $148$ &$1.38$ & $33$ & $1.38$\\
			$C$, J/(K kg) & $713$ & $746$ &$306$ & $746$ \\
			$\Phi$ & $5\times 10^{-5}$ & $4\times 10^{-10}$ & $2\times 10^{-4}$ & $4\times 10^{-5}$ \\
			$n$ & 3.48 & 1.45 & 2.06 & 1.45\\
			\hline\hline
		\end{tabular}
		\label{tab:Material}
\end{table}

\begin{table}[!htbp]
	\centering
		\caption{Material properties of the meta-etalon components at 124 K \cite{heinert2014database}.}
		\begin{tabular}{|c|c|c|c|c|}
		\hline\hline
			Parameter & Silicon & Substrate & Ta$_2$O$_5$ layer & SiO$_2$ layer \\
			\hline\hline
			$\beta$, 1/K & $9\times 10^{-5}$ & $4.2\times 10^{-6}$ & $1.4\times 10^{-5}$ & $4.2\times 10^{-6}$ \\
			$\alpha$, 1/K & $\approx 0$ & $-4.8\times 10^{-7}$ & $3.6\times 10^{-6}$ & $-4.8\times 10^{-7}$ \\
			$\rho$, kg/m$^3$ & $2331$ & $2203$ & $6850$ & $2202$ \\
			$Y$, Pa & $130\times 10^9$ & $72\times 10^9$ & $140\times 10^9$ & $72\times 10^9$ \\
			$\nu$ & $0.28$ & $0.16$ & $0.23$ & $0.16$ \\
			$\kappa$, W/(K m) & $638$ &$0.804$ & $33$ & $0.804$\\
			$C$, J/(K kg) & $328$ & $339$ &$306$ & $339$\\
			$\Phi$ & $5\times 10^{-6}$ & $10^{-7}$ & $2\times 10^{-4}$ & $4\times 10^{-5}$\\
			$n$ & 3.48 & 1.45 & 2.06 & 1.45\\
			\hline\hline
		\end{tabular}
		\label{tab:Material2}
\end{table}

\cleardoublepage

\section*{References}
\bibliography{References_Etalon}

\end{document}